\documentclass[10pt,twocolumn]{article}

\usepackage{authblk}
\usepackage{amsmath,amssymb}
\usepackage{graphicx}
\usepackage{hyperref}
\usepackage{cite}
\hypersetup{
    colorlinks=true,
    linkcolor=blue,
    citecolor=blue,
    urlcolor=blue
}

\title{Encoding Hyperuniformity in Geometric Networks:
Structural Signatures and Diffusion Dynamics}

\author[a]{Yongyi Wang}
\author[b]{Eli Newby}
\author[c,d]{Yang Jiao}
\author[e,f,g,h]{Salvatore Torquato}
\author[a]{R{\'e}ka Albert\textsuperscript{*}}
\affil[a]{Department of Physics, Pennsylvania State University,
University Park, PA 16802}

\affil[b]{Department of Cancer Sciences,
Cleveland Clinic Research, Cleveland, OH 44106}

\affil[c]{Materials Science and Engineering,
Arizona State University, Tempe, AZ 85287}

\affil[d]{Department of Physics,
Arizona State University, Tempe, AZ 85287}

\affil[e]{Department of Chemistry,
Princeton University, Princeton, NJ 08544}

\affil[f]{Department of Physics,
Princeton University, Princeton, NJ 08544}

\affil[g]{Princeton Institute of Materials,
Princeton University, Princeton, NJ 08544}

\affil[h]{Program in Applied and Computational Mathematics,
Princeton University, Princeton, NJ 08544}

\date{}

\begin{document}

\twocolumn[{%
\begin{@twocolumnfalse}
\maketitle
\begin{abstract}
Disordered hyperuniformity is a recently discovered exotic state exhibiting anomalously suppressed density fluctuations at long wavelengths, and its generalization to network systems remains an important open question. Here, we show that, across a wide range of systems, geometric networks retain substantially more long-wavelength density information from the underlying point configurations than Delaunay triangulations. Within geometric networks, hyperuniformity manifests as suppressed vertex-to-vertex fluctuations in a range of network structural observables, including closeness centrality, vertex strength, and eigenvector centrality. These measures not only distinguish hyperuniform networks from non-hyperuniform ones, but also separate near-hyperuniform configurations from hyperuniform systems with strong local disorder, solving a central challenge in network hyperuniformity. We connect these structural signatures to network dynamics by introducing a relaxation-time centrality based on diffusion and demonstrate that hyperuniform geometric networks exhibit more homogeneous relaxation behavior than non-hyperuniform geometric networks. These results reveal how the suppression of large-scale density fluctuations manifests in both network structure and diffusion dynamics.
\end{abstract}

\noindent\textbf{Keywords:}
Network hyperuniformity;
geometric networks;
diffusion dynamics
\par\vspace{2em}
\end{@twocolumnfalse}
}]
\begingroup
\renewcommand{\thefootnote}{\fnsymbol{footnote}}
\footnotetext[1]{Corresponding author:
\href{mailto:rza1@psu.edu}{rza1@psu.edu}}
\endgroup
\section{Introduction}
Disordered hyperuniform (DHU) many-body systems possess a ``hidden order'' manifested as a complete suppression of normalized infinite-wavelength density fluctuations in a manner similar to crystals. Yet, they lack conventional long-range order, possess no broken symmetry and retain the structural isotropy of amorphous materials \cite{To03, To18a}. These unique structural characteristics give rise to distinctive physical properties, including complete isotropic photonic \cite{Fl09, klatt2022wave} and phononic band gaps \cite{Gkantzounis2017, Valier-Brasier2022, RomeroGarcia2021}, nearly optimal transport properties \cite{Zh16, spreadability,jiao2026acoustic} and mechanical response \cite{Xu17, ref36}, and stealthy propagation of electromagnetic waves \cite{Le16, kim2023effective, Torquato2021_EM}. DHU is now recognized as a widespread state of matter \cite{To18a, chen2023disordered, lei2024non}, appearing in equilibrium systems at positive temperature $T$ \cite{WangTorquato2024} and their ground states \cite{Ba09, Torquato2015}, driven nonequilibrium and active systems \cite{zheng2024universal, Za11a, He15, Le19, lei2019hydrodynamics}, cosmological models \cite{Ga02}, random-organizing systems \cite{hexner2017noise}, and quantum many-body systems \cite{Wang_Samajdar2024, Vanoni_Karcher2026, Ge19quantum, sakai2022quantum, Zh20, Ch21, chen2025anomalous}. More recently, disordered hyperuniform organization has also been identified in biological systems spanning a wide range of length scales \cite{Ji14, Ma15, tang2024tunable, li2025fluidization, leaf_network, ge2023hidden, hu2025causes}. 

Hyperuniformity is characterized by a local number variance $\sigma_N^2(R)$ associated with a spherical window of radius $R$ in $\mathbb{R}^d$ that grows more slowly than the window volume in the large-$R$ limit \cite{To03, To18a}, i.e., 
\begin{equation}
\lim_{R\rightarrow \infty} \frac{\sigma_N^2(R)}{R^d} = 0. 
\end{equation}
Equivalently, the static structure factor $S({\bf k})$ vanishes in the infinite-wavelength (or zero-wavenumber) limit, i.e., 
\begin{equation}
    \lim_{|{\bf k}|\rightarrow 0}S({\bf k}) = 0
\end{equation}
where ${\bf k}$ is the wavevector. For statistically isotropic systems, the structure factor only depends on the wavenumber, $k \equiv |{\bf k}|$. Its small-$k$ scaling behavior, $S(k) \sim k^{\alpha}$ with $\alpha>0$ for hyperuniform systems, determines the large-$R$ asymptotic behavior of $\sigma_N^2(R)$. Furthermore, all hyperuniform systems can be categorized into three different classes: $\sigma_N^2(R) \sim R^{d-1}$ for $\alpha>1$ (class I); $\sigma_N^2(R)
\sim R^{d-1}\ln(R)$ for $\alpha=1$ (class II); and $\sigma_N^2(R)\sim R^{d-\alpha}$ for $0<\alpha<1$ (class III). Systems for which $S(k)$ approaches a bounded positive constant as $k\rightarrow0$ are standard non-hyperuniform systems, whereas antihyperuniform systems exhibit enhanced long-wavelength density fluctuations reflected by a divergent structure factor, i.e., $\lim_{k\rightarrow0}S(k)\rightarrow+\infty$.

Beyond continuum descriptions, it is important to understand how hyperuniformity manifests in networked systems, where structure is encoded through vertices and edges. This question is particularly relevant because many physical, biological, and engineered systems are naturally described as discrete interacting units \cite{Networks_Newman, Rev_network2, Rev_network3, epi_net1, network_material}. Indeed, hyperuniform network architectures have already attracted interest in settings ranging from cellular materials \cite{ref30, Zh16, ref36, tang2024tunable} and disordered networks with unusual thermal or acoustic behavior \cite{jiao2026acoustic} to biological systems such as looped leaf-vein networks \cite{leaf_network} and active matter systems such as magnetic robotic spinner networks \cite{spinner_network}. These developments suggest that a network perspective provides a natural framework for investigating how the suppressed large-scale fluctuations characteristic of DHU structures influence system-level organization and functionality.

Despite recent progress, a central challenge remains: commonly used Delaunay triangulation and Voronoi tessellation networks can often distinguish hyperuniform systems from uncorrelated ones, but they struggle to distinguish highly locally ordered near-hyperuniform systems from DHU systems with strong local disorder \cite{Eli_PRE, HU_Eli_2}. Recent work by Newby et al. showed that the density information encoded by Delaunay triangulation networks is only weakly correlated with that of the underlying hyperuniform point configurations at long wavelengths \cite{HU_Eli_2}. These limitations motivate us to study network representations that retain more long-range density information than Delaunay-based local constructions.

A broad range of spatial network constructions can be considered, including $k$-nearest-neighbor, Gabriel, and relative-neighborhood graphs. Notably, constructions that are subgraphs of the Delaunay triangulation, including the Gabriel and relative-neighborhood graphs, cannot reduce the shortest-path distortion relative to Euclidean distances below that of the Delaunay triangulation \cite{Gabriel_RNG}. To systematically probe the retention of long-wavelength density information, we therefore seek a network representation whose spatial range of connectivity can be controlled directly by a physical length scale.

For this purpose, we consider geometric networks, in which edges are assigned between all pairs of points within a finite connection radius $R$. Increasing $R$ tunes the network from predominantly local connectivity toward the complete-graph limit. Using this construction, we study representative 2D point configurations spanning antihyperuniform, standard non-hyperuniform, near-hyperuniform, and hyperuniform systems. We show that geometric networks retain substantially more long-wavelength density information from the underlying point configurations than Delaunay triangulations across the systems considered. 

Importantly, we demonstrate that signatures of hyperuniformity manifest in graph-level observables. Specifically, hyperuniform geometric networks exhibit suppressed vertex-to-vertex fluctuations in closeness centrality, vertex strength, and eigenvector centrality, allowing near-hyperuniform systems with strong local order to be distinguished from hyperuniform systems with strong local disorder. We further introduce a relaxation-time centrality based on diffusion dynamics and show that hyperuniform geometric networks exhibit more homogeneous relaxation behavior than their non-hyperuniform counterparts. Together, these results reveal the structural and dynamical signatures of hyperuniform geometric networks.

\section{Results}
\subsection*{High density correlation of geometric networks. }
\begin{figure*}[!t]
\centering
\includegraphics[scale=0.5]{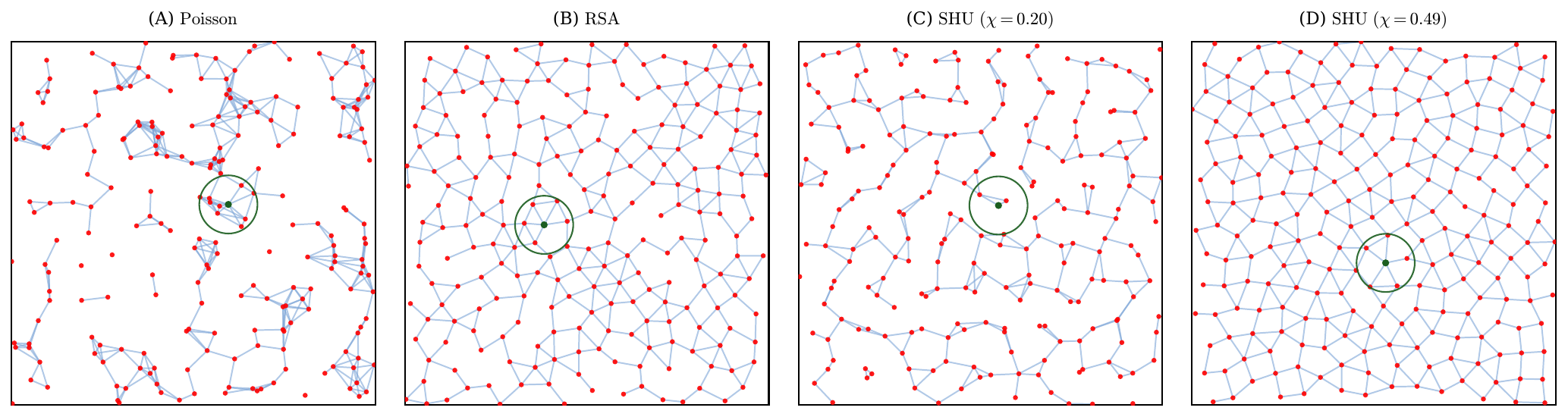}
\caption{Demonstration of geometric networks with $R=1.2$ ($0.012L$) constructed from (A) Poisson, (B) RSA, (C) SHU ($\chi=0.20$), and (D) SHU ($\chi=0.49$) point configurations. Each point is connected to all other points within the connection radius $R$, as illustrated by the green circle. Although the near-hyperuniform RSA configuration exhibits stronger local order than the SHU ($\chi=0.20$) disordered hyperuniform systems, the latter possesses hidden long-range order. Therefore, distinguishing these systems requires a network construction that is sensitive to structural information beyond short-range local order.}
\label{fig: config_GN}
\end{figure*}

We derive geometric networks from a broad spectrum of disordered point configurations within an $L \times L$ square domain, including three stealthy hyperuniform (SHU) systems of increasing degree of local order described by a stealthiness parameter $\chi$ \cite{Batten2008}, one-component-plasma (OCP) systems \cite{Wang2023, WangTorquato2024}, near-hyperuniform random-sequential-addition (RSA) configurations, standard non-hyperuniform Poisson configurations, and antihyperuniform hyperplane-intersection-process (HIP) configurations \cite{Stoyan1995}; see \textit{Materials and Methods}. Unless otherwise specified, all calculations are performed for systems with $L=100$. An illustration of geometric network construction is shown in Fig.~\ref{fig: config_GN} for representative point configurations.

\begin{figure*}[!t]
\centering
\includegraphics[scale=0.5]{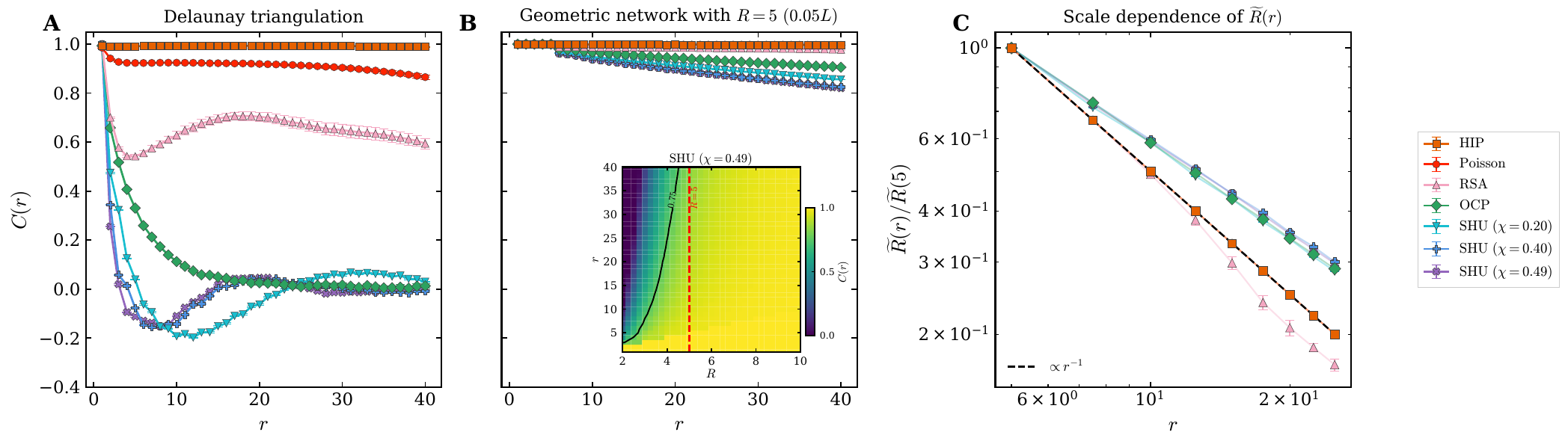}
\caption{Network density correlation $C(r)$, as defined in Eq.~\ref{eqn: C(r)}, for (A) Delaunay triangulation networks and (B) geometric networks with connection radius $R=5$ ($0.05L$), constructed from the point configurations considered here. The inset in (B) shows $C(r)$ for the SHU ($\chi=0.49$) system as a function of spatial scale $r$ over a range of connection radii $R$. The black contour denotes $C_0=0.75$, and the red dashed line marks $R=5$. (C) Scale dependence of the relative connection range $\widetilde{R}(r)$, normalized by $\widetilde{R}(5)$, as defined in Eq.~\ref{eqn: Rc def}. For all systems, $\widetilde{R}(r)$ decreases with $r$, indicating that $R_c(r)$ grows more slowly than the spatial scale being probed over the scales examined.} 
\label{fig:density-correlation}
\end{figure*}

The network density correlation $C(r)$ quantifies the extent to which a derived network retains the local density information of the underlying point configuration by measuring the correlation between Euclidean and graph-based neighborhood densities \cite{HU_Eli_2}; see \textit{Materials and Methods}. Figure \ref{fig:density-correlation} shows the network density correlations of geometric networks at a representative connection radius $R=5$ ($0.05L$), along with those of Delaunay triangulations for comparison. At sufficiently short length scales, graph-based and Euclidean neighborhoods coincide because nearby points are directly connected, giving $C(r)\approx1$. At larger length scales, shortest-path distortion causes graph-based neighborhoods to increasingly deviate from their Euclidean counterparts, generally reducing $C(r)$. Nonetheless, the two network constructions exhibit markedly different behaviors as larger length scales are probed.

As shown in Fig.~\ref{fig:density-correlation}, Delaunay triangulation retains density information reasonably well for the non-hyperuniform HIP and Poisson systems, and to a lesser extent for the near-hyperuniform RSA system. By contrast, for hyperuniform systems, $C(r)$ decays rapidly with increasing $r$, indicating that Delaunay networks are not or only weakly correlated with the underlying point configurations at larger length scales. Thus, Delaunay triangulation does not retain information on the long-wavelength density fluctuations of hyperuniform systems, consistent with Ref.~\cite{HU_Eli_2}.

By contrast, geometric networks exhibit substantially stronger density correlations at large $r$ for the hyperuniform systems, as evidenced by the markedly higher $C(r)$ values in Fig.~\ref{fig:density-correlation}B compared with the corresponding Delaunay networks in Fig.~\ref{fig:density-correlation}A. Among the systems considered, the high-$\chi$ SHU configurations exhibit the lowest $C(r)$ at a given $r$. The inset of Fig.~\ref{fig:density-correlation}B shows the dependence of $C(r)$ on the connection radius $R$ for the $\chi=0.49$ SHU system, indicating that high density correlations can be achieved with connection radii that remain small compared with the system size. However, for any fixed $R$, $C(r)$ generally decreases as increasingly large length scales $r$ are probed. Maintaining a prescribed level of density correlation at larger $r$ therefore requires a corresponding increase in the connection radius. This raises a natural question: how does the connection radius required to retain density information scale with the spatial scale being represented?

We address this question by defining a critical connection radius $R_c(r)$ as the minimum $R$ required to achieve a prescribed density correlation $C(r)\ge C_0$, with $C_0=3/4$, while restricting the analysis to connected networks. We then define the relative connection range
\begin{equation}\label{eqn: Rc def}
    \widetilde{R}(r)=\frac{R_c(r)}{r},
\end{equation}
which measures the connection radius required to retain density information relative to the spatial scale being probed. An increasing $\widetilde{R}(r)$ would indicate that the connection radius required to retain density information constitutes an increasing fraction of the spatial scale being represented. The resulting network would therefore become progressively denser and less local as larger length scales are probed. Conversely, a decreasing $\widetilde{R}(r)$ indicates that, even if $R_c(r)$ may increase with $r$, it does so more slowly than the probed scale, so that the required connections span a progressively smaller fraction of that scale.

Figure~\ref{fig:density-correlation}C shows that $\widetilde{R}(r)$ decreases with $r$ for all systems considered. For the antihyperuniform HIP and nonhyperuniform Poisson systems, the threshold $C(r)\ge C_0$ is already satisfied once the network becomes connected. Consequently, $R_c(r)$ is set by the connectivity threshold and remains approximately constant, giving $\widetilde{R}(r)\propto r^{-1}$. 

For the hyperuniform systems considered, $R_c(r)$ increases with $r$, but more slowly than the probed length scale, so that $\widetilde{R}(r)$ is still a decreasing function. Interestingly, $R_c(r)$ is well described over the scales examined by a sublinear power law, $R_c(r)\sim r^{\gamma}$ with $0<\gamma<1$, corresponding to $\widetilde{R}(r)\sim r^{\beta}$ with $-1<\beta=\gamma-1<0$, as shown in Figure~\ref{fig:density-correlation}C. Within the SHU family, the fitted exponent depends only weakly on the stealthiness parameter $\chi$, and representative finite-size analyses for $\chi=0.20$ and $0.49$ show consistent fitted exponents $\gamma$ across several system sizes; see \textit{Supporting Information}. Thus, geometric networks can retain long-wavelength density information while the required connection range remains increasingly local relative to the spatial scale being probed.
\subsection*{Suppressed centrality fluctuations reveal hidden hyperuniform order}
We next investigate whether the suppression of density fluctuations in hyperuniform point configurations is reflected in reduced structural heterogeneity of the networks. We focus on three structural observables, namely closeness centrality, vertex strength, and eigenvector centrality, which probe different aspects of the spatial organization surrounding each vertex. When the network preserves information about the underlying point configuration over sufficiently large length scales, fluctuations in these observables can reflect fluctuations in the underlying density field. The suppression of long-wavelength density fluctuations in hyperuniform systems may therefore give rise to reduced heterogeneity in these network measures.

\begin{figure*}[!t]
\centering
\includegraphics[scale=0.45]{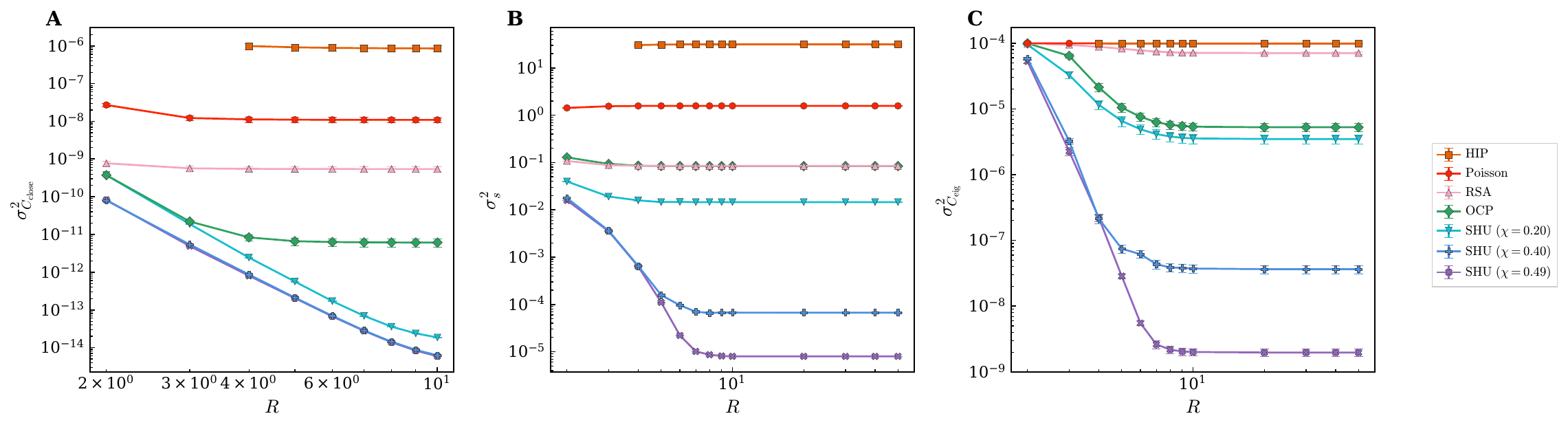}
\caption{Variance of (A) the closeness centrality, $\sigma^2_{C_{\mathrm{close}}}$, (B) the vertex strength, $\sigma_{s}^2$, and (C) the eigenvector centrality, $\sigma^2_{C_{\mathrm{eig}}}$, for geometric networks as a function of the connection radius $R$. Hyperuniform geometric networks exhibit suppressed variance in all three structural centrality measures. The hierarchy of the variances reflects the degree of hyperuniformity of the underlying point configurations. We note that the numerical values of different centrality measures are not directly comparable.}
\label{fig: centralities}
\end{figure*}

The closeness centrality $C_{\mathrm{close}}(i)$ quantifies the global accessibility of a vertex through its average shortest-path distance to the rest of the network; see \textit{Materials and Methods}. When graph neighborhoods closely approximate their Euclidean counterparts, fluctuations in the number of vertices within a given graph distance reflect density fluctuations in the underlying point configuration. Suppression of these density fluctuations over large scales can reduce vertex-to-vertex variations in closeness centrality. To quantify this effect, we compute the variance of closeness centrality across the network. Figure~\ref{fig: centralities}A shows $\sigma^2_{C_{\mathrm{close}}}$ as a function of the connection radius $R$ for all point configurations considered. 

We find that geometric networks constructed from hyperuniform configurations exhibit smaller closeness variances than their non-hyperuniform counterparts. Importantly, the variances reflect the degree of hyperuniformity of the underlying point configurations. This ordering persists throughout the range of connection radii $R$ considered.

While different systems are clearly separated at larger $R$, the differences in closeness variance between the near-hyperuniform RSA system and the hyperuniform OCP and SHU ($\chi=0.20$) systems start to diminish at small $R$. At these connection radii, geometric networks retain less long-range density information, reducing their sensitivity to distinctions that emerge primarily at long wavelengths. This also helps explain the insufficient sensitivity of Delaunay triangulation, whose predominantly local connectivity similarly limits the density information retained from extended spatial neighborhoods. 

We next examine whether the suppression of structural heterogeneity also appears in interaction-based network measures that probe spatial organization at different levels. Vertex strength depends only on the weighted connections to directly connected neighbors and therefore probes the immediate network neighborhood. By contrast, eigenvector centrality recursively incorporates the connectivity of neighboring vertices and thus contains information from progressively more distant parts of the network \cite{Networks_Newman}. We therefore use these two measures to probe structural heterogeneity at complementary network scales.

We define an edge weight $W_{ij}$ that decays with the Euclidean distance, i.e., 
\begin{equation}\label{eqn: weight assignment}
    W_{ij} =
    \begin{cases}
    \exp\left(\frac{-d_{ij}}{\lambda}\right), & d_{ij} < R, \\
    0, & d_{ij} \ge R,
    \end{cases}
\end{equation}
where \(\lambda\) is a characteristic interaction length scale. This functional form is bounded and avoids the short-distance divergence associated with inverse-distance weights such as $W_{ij}=1/d_{ij}$. Since the number density is normalized to unity, we take $\lambda=1$, so that the interaction length scale is comparable to the characteristic interparticle spacing; see \textit{Materials and Methods}.

Figure~\ref{fig: centralities}B and C show the variances of the vertex strength and eigenvector centrality, respectively, as functions of the connection radius $R$. As with closeness centrality, hyperuniform geometric networks exhibit systematically smaller variances than their non-hyperuniform counterparts once $R\gtrsim5$ for $\sigma^2_s$ and $R\gtrsim7$ for $\sigma^2_{C_{\mathrm{eig}}}$. Thus, suppressed structural heterogeneity associated with hyperuniformity appears across network measures that probe the spatial organization at different network scales. In particular, these measures distinguish the near-hyperuniform RSA system from the hyperuniform OCP and SHU ($\chi=0.20$) systems despite the stronger local order of RSA.

\subsection*{Hyperuniformity suppresses diffusion-relaxation heterogeneity}
To connect the structural signatures of hyperuniformity to transport properties, we study diffusion relaxation on geometric networks by considering the spreading of a localized perturbation. The relaxation of a localized perturbation depends on how the initial vertex is connected to the rest of the network. Structural bottlenecks and weakly connected regions can lead to slowly relaxing modes, causing perturbations initiated in such regions to approach equilibrium more slowly \cite{Networks_Newman}. By contrast, perturbations to vertices that are well connected to the broader network tend to relax more rapidly. Vertex-to-vertex variations in relaxation time therefore provide a dynamical measure of heterogeneity in network accessibility.

For each vertex $i$, we define the relaxation-time centrality $C_{\mathrm{heat}}(i)$ as the inverse mean relaxation time for diffusion initialized at $i$ to approach equilibrium, 
\begin{equation}\label{eqn: C_heat main}
    C_{\mathrm{heat}}(i)= 2\frac{\sum_{k=2}^N\frac{1}{\lambda_k}\phi^2_k(i)}{\sum_{k=2}^N\frac{1}{\lambda_k^2}\phi^2_k(i)} = \frac{2\left(L^{+}\right)_{ii}}{\left[\left(L^{+}\right)^2\right]_{ii}},
\end{equation}
where $\lambda_k$ and $\phi_k$ are the eigenvalues and corresponding eigenvectors of the graph Laplacian $L_{ij}=\delta_{ij}\sum_{k}W_{ik}-W_{ij}$ in ascending order ($\lambda_1=0<\lambda_2\leq\cdots\leq\lambda_N$), and $L^{+}$ is the Moore--Penrose pseudoinverse of the Laplacian matrix; see \textit{Materials and Methods}. The heterogeneity of the diffusion relaxation time is then characterized by the variance $\sigma^2_{C_{\mathrm{heat}}}$ across vertices. 

\begin{figure} 
\centering
\includegraphics[width=0.4\textwidth]{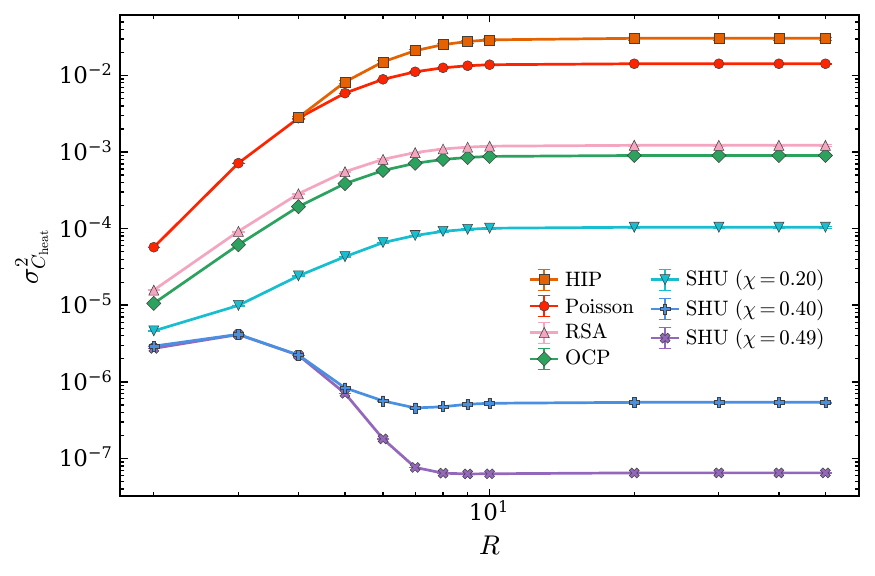}
\caption{Variance of the relaxation-time centrality, $\sigma^2_{C_{\mathrm{heat}}}$, for geometric networks as a function of the connection radius $R$. Similar to the structural descriptors, diffusion dynamics capture the hyperuniform character of the underlying point configurations. Across the broad spectrum of systems considered, hyperuniform geometric networks exhibit reduced diffusion-relaxation heterogeneity.}
\label{fig: heat GN scaling}
\end{figure}

Figure \ref{fig: heat GN scaling} shows $\sigma^2_{C_{\mathrm{heat}}}$ across a range of connection radii $R$ that yield connected geometric networks. We find that $\sigma^2_{C_{\mathrm{heat}}}$ depends strongly on the degree of hyperuniformity of the underlying point configurations. In hyperuniform networks, vertices exhibit nearly identical relaxation times $\overline{T}(i)$, indicating highly uniform spreading dynamics despite structural disorder. The reduced variance therefore indicates that diffusion relaxation is less sensitive to the vertex at which the perturbation is initially introduced. This enhanced uniformity of relaxation is qualitatively consistent with previous studies of diffusion spreadability in hyperuniform continuum media \cite{spreadability}.

\begin{figure*}
\centering
\includegraphics[scale=0.45]{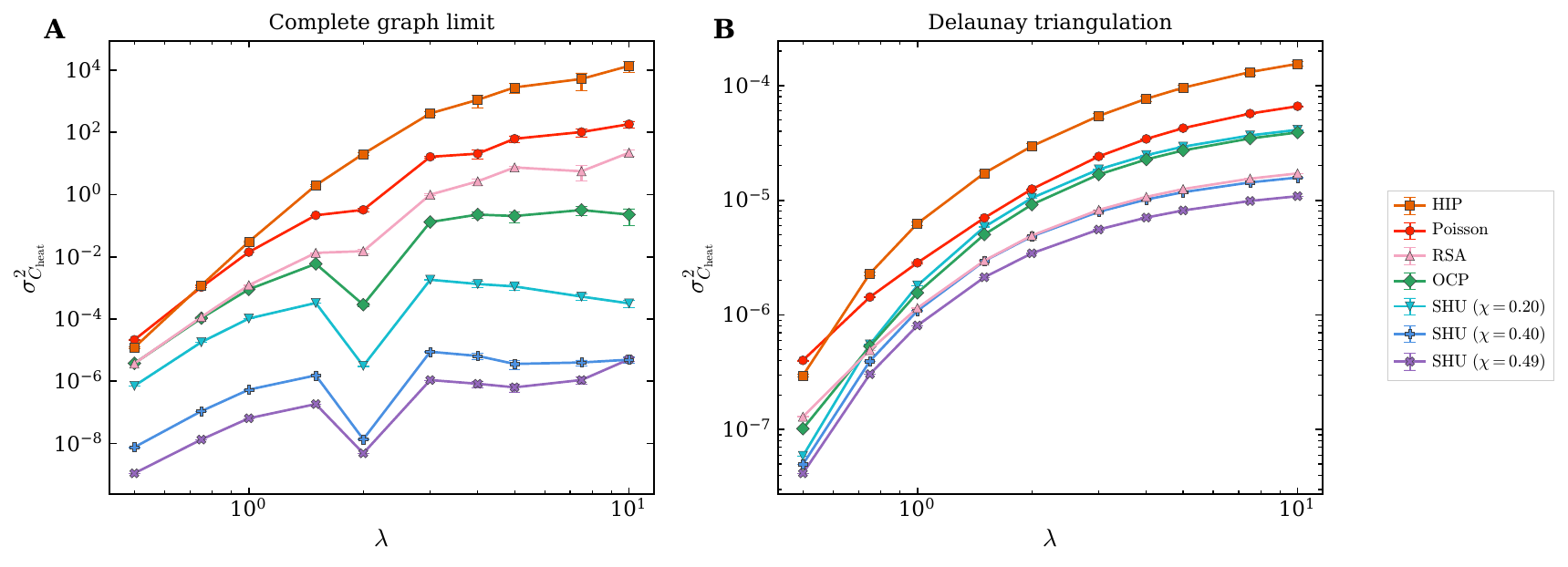}
\caption{Variance of the relaxation-time centrality, $\sigma^2_{C_{\mathrm{heat}}}$, for (A) weighted complete graphs and (B) Delaunay triangulation networks over a wide range of $\lambda$. Because geometric networks converge rapidly toward the complete-graph limit, panel A shows that the diffusion hierarchy persists in the weighted complete-graph limit. By contrast, Delaunay triangulation cannot distinguish the near-hyperuniform RSA configurations from the more locally disordered OCP and SHU $\chi=0.20$ systems.}
\label{fig: C_heat scaling}
\end{figure*}

We next examine the robustness of this diffusion hierarchy with respect to the connection radius $R$ and interaction length scale $\lambda$. As the connection radius increases, $\sigma^2_{C_{\mathrm{heat}}}$ rapidly approaches the weighted complete-graph limit since interactions with $d_{ij}\gg\lambda$ contribute negligibly to the diffusion dynamics. Figure~\ref{fig: C_heat scaling}A further shows that the observed ordering of $\sigma^2_{C_{\mathrm{heat}}}$ persists over a broad range of $\lambda$, except in the $\lambda\ll1$ regime where interactions become strongly localized and no longer probe the collective spatial organization of the point configurations.

On the other hand, Delaunay triangulation does not clearly distinguish the relaxation behavior of the near-hyperuniform RSA system from the hyperuniform OCP and SHU ($\chi=0.20$) systems across the range of $\lambda$ considered, as shown in Figure~\ref{fig: C_heat scaling}B. This behavior corroborates the known limitations of Delaunay-based constructions in capturing differences that arise primarily at long wavelengths~\cite{Eli_PRE, HU_Eli_2}.

\section{Conclusion and Discussion}
In this work, we addressed a central challenge in the study of network hyperuniformity: how long-wavelength density information in point configurations can be represented in spatial networks. Across the broad range of systems considered, geometric networks retain substantially more long-wavelength density information than Delaunay triangulations while the connectivity remains finite. At the graph level, hyperuniform geometric networks exhibit suppressed vertex-to-vertex fluctuations in closeness centrality, vertex strength, and eigenvector centrality. Importantly, these measures distinguish the near-hyperuniform RSA system from the more locally disordered hyperuniform OCP and SHU ($\chi=0.20$) systems, a distinction that could not be achieved by Delaunay triangulations. These results establish geometric networks as a tunable framework for studying how long-wavelength density fluctuations are reflected in network structure.

Beyond structural characterization, hyperuniformity also manifests in diffusion dynamics on geometric networks. Using the proposed relaxation-time centrality $C_{\mathrm{heat}}$, we find that localized perturbations relax more uniformly in hyperuniform networks than in non-hyperuniform networks. Diffusive relaxation is therefore less sensitive to the vertex at which the perturbation is initially introduced in hyperuniform networks. Hence, geometric networks reveal the signatures of hyperuniformity in both structural and dynamical properties. 

The tunable connection radius also provides a connection between finite-range geometric networks and weighted complete graphs. For the weighted observables considered here, increasing $R$ leads rapidly toward the weighted complete-graph limit. This suggests a possible route for relating finite-radius geometric networks to analytical results obtained for weighted complete-graph constructions. For example, recent work by Jiao has established an analytical relationship between the scaling of the heat capacity of elastic networks generated by smooth radial kernels and the small-$k$ behavior of the structure factor of the underlying point configurations. Determining the conditions under which such analytical relationships can be extended to finite-radius geometric networks remains an interesting open question.

More broadly, our results demonstrate that the hidden long-wavelength order of hyperuniform point configurations can be expressed directly through structural and dynamical properties of spatial networks. The suppression of graph-level fluctuations demonstrated here suggests that hyperuniformity may provide a useful organizing principle for designing networks with enhanced structural and dynamical uniformity. In future work, we aim to further connect the properties of hyperuniform networks and their underlying point statistics through analytical approaches.
\section{Materials and Methods}
\subsection*{Generation of the point configurations}
The HIP configurations result from the intersection points of random, independent placements of hyperplanes, possessing hyper-fluctuations \cite{Stoyan1995}. The Poisson configurations are generated by randomly placed points in a prescribed domain, leading to completely uncorrelated distributions. The RSA process generates a saturated packing by sequentially placing fixed-radius hard spheres at random nonoverlapping positions within the prescribed domain until no further insertions are possible \cite{Zhang2013}. The {\it nonstealthy hyperuniform} OCP point process is generated using a Gaussian pair statistic, with $g_2(r) = 1-exp(-\pi r^2)$, which is a Class-I hyperuniform system, with $S(k) \sim k^2$ for small $k$ values \cite{Wang2023, WangTorquato2024}. 
Stealthy hyperuniform (SHU) systems are a special subset of class-I hyperuniform systems possessing a zero structure factor for a range of wavevectors around the origin, i.e.,
\begin{equation}
S({\bf k}) = 0, \quad\text{for}\quad {\bf k} \in \Omega,
\label{eq_stealthy}
\end{equation}
excluding the forward scattering. SHU systems are determined by a tuning parameter $\chi$ measuring the fraction of the independently constrained degrees of freedom $M$ within the exclusion region $\Omega$ \cite{Torquato2015}, i.e., $  \chi = {M}/{(N-1)d}$, 
where $N$ is the total number of points in the systems. 
The disordered SHU configurations are generated using the ``collective coordinates'' approach \cite{Torquato2015, Batten2008}.

For all aforementioned systems, we generate $N=L^2$ points in an $L \times L$ square domain with periodic boundary conditions so that the number density is normalized to unity, i.e., $\rho=1$. Throughout this work, we use $L=100$ for all systems considered. For the hyperuniform configurations, we use 20 realizations of disordered SHU systems with $\chi=0.20$, $0.40$, and $0.49$, and 9 realizations of OCP systems. For comparison, we use 10 realizations of RSA configurations, 20 realizations of Poisson configurations, and 10 realizations of HIP configurations.
\subsection*{Network density correlation} 
The normalized local density fluctuation of the point configuration at point $i$ is defined as: 
\begin{equation}
\delta\rho_P(i, r)=\frac{N_P(i,r)-\langle N_P(r) \rangle}{\sigma_P(r)},
\end{equation}
where $N_P(i,r)$ is the number of points within a circle of radius $r$ centered at point $i$, $\langle N_P(r) \rangle$ is the average of $N_P(i,r)$ over all points, and $\sigma_P(r)$ is the corresponding standard deviation. Its graph-theoretic counterpart is defined analogously using the graph distance:
\begin{equation}
\delta\rho_G(i, r)=\frac{N_G(i,r)-\langle N_G(r) \rangle}{\sigma_G(r)},
\end{equation}
A network preserves the local density information of its underlying point configuration at scale $r$ if $N_G(i,r)$ is strongly correlated with $N_P(i,r)$ across vertices. In the limiting case of a complete graph, $N_G(i,r)=N_P(i,r)$ since no information is lost. We define the network density correlation as
\begin{equation}\label{eqn: C(r)}
C(r)=\frac{1}{N-1}\sum_{i=1}^{N}\delta\rho_P(i,r)\delta\rho_G(i,r)
\end{equation}
which is the Pearson correlation coefficient between $\{N_G(i,r)\}_{i=1,\cdots,N}$ and $\{N_P(i,r)\}_{i=1,\cdots,N}$ where $N$ is the number of points in the configuration, or equivalently the number of vertices in the network. 
\subsection*{Closeness centrality}
Closeness centrality measures the inverse average shortest distance from a vertex to all other vertices in the network. The edges are weighted by the Euclidean distance between the corresponding vertices. For a vertex $i$, closeness centrality is defined as \cite{Networks_Newman}: 
\begin{equation}\label{eqn: closeness}
    C_{\mathrm{close}}(i) = \left(\frac{1}{N-1}\sum_{j\neq i}d_{ij}\right)^{-1}
\end{equation}
where $N$ is the total number of vertices in the connected network and $d_{ij}$ is the shortest-path distance between vertex $i$ and $j$.
\subsection*{Vertex strength}
The strength of a vertex is defined as
\begin{equation}
    s_i = \sum_{j} W_{ij},
\end{equation}
where $W_{ij}$ is defined in Equation \ref{eqn: weight assignment}.
\subsection*{Eigenvector centrality}
The eigenvector centrality is obtained from \cite{Networks_Newman}
\begin{equation}
    W \vec{x} = \lambda_{\max} \vec{x},
\end{equation}
where $W$ is the matrix of edge weights $W_{ij}$, $\lambda_{\max}$ is the largest eigenvalue of $W$ and $\vec{x}$ is the corresponding eigenvector, with $C_{\text{eig}}(i) = x_i$. 
\subsection*{Diffusion relaxation time}
The diffusion equation on networks is \cite{diffusion_ref}
\begin{equation}
    \frac{d\vec{u}}{dt}=-L\vec{u}
\end{equation}
where $L=D-W$ is the weighted Laplacian matrix of the network with $D_{ij}=\delta_{ij}\sum_{k}W_{ik}$ and $W_{ij}$ is defined in Eq.~\ref{eqn: weight assignment}. The diffusion equation has a solution 
\begin{equation}
    \vec{u}(t)=e^{-tL}\vec{u}(0)
\end{equation}
with the initial condition $\vec{u}(0)$. To quantify the diffusive properties of a given vertex $i$, we study the time scale for the system to approach equilibrium when diffusion starts from $i$, i.e.,  $\vec{u}(0)=\hat{e}_i$. The deviation from the steady-state solution is
\begin{equation}\label{eqn: h_i(t)}
    h_i(t) =\left(e^{-tL}-\Pi\right)\hat{e}_i= \vec{u}(t)-\vec{u}(\infty)
\end{equation}
where $\Pi= \frac{1}{N}\vec{\mathbf{1}}\vec{\mathbf{1}}^T$ is the steady-state projector with $\vec{\mathbf{1}}$ being the all-one vector. We sum over the squared deviations from the equilibrium state of all vertices, and define the energy of the system to be 
\begin{equation}\label{eqn: E_i(t)}
    E_i(t)=||h_i(t)||^2=\sum_{j}h_{ji}(t)^2
\end{equation}
The \textit{mean relaxation time} for the initial condition $u(0)=\hat{e}_i$ is therefore, 
\begin{equation}\label{eqn: relaxation time}
    \bar{T}(i) = \frac{\int_0^{\infty}tE_i(t)dt}{\int_0^{\infty}E_i(t)dt}
\end{equation}
We define the relaxation-time centrality $C_{\mathrm{heat}}$ as
\begin{equation}
    C_{\mathrm{heat}}(i)\equiv\frac{1}{\bar{T}(i)}= \frac{2\left(L^{+}\right)_{ii}}{\left(\left(L^{+}\right)^2\right)_{ii}} 
\end{equation}
where the second equation comes from spectral decomposition; see \textit{Supporting Information}.
\section*{Data Availability}
The raw data supporting the findings of this study are available at \cite{Wang2026GN}.
\section*{Acknowledgments}
This work was supported by the Army Research Office under Cooperative Agreement Number W911NF-22-2-0103.
\begingroup
\small
\bibliographystyle{unsrt}
\bibliography{apssamp,network}
\endgroup
\clearpage
\onecolumn
\section*{Supporting Information}
\subsection*{Diffusion relaxation time}
The rate change in the concentration at vertex $i$, $u_i$, due to the presence of another vertex $j$ is 
\begin{equation}
    \dot{u}_i = W_{ij}(u_j-u_i)
\end{equation}
Summing over all neighboring vertices gives the diffusion equation on networks \cite{diffusion_ref}
\begin{equation}
    \frac{d\vec{u}}{dt}=-L\vec{u}
\end{equation}
where $L=D-W$ is the weighted Laplacian matrix of the network with $D_{ij}=\delta_{ij}\sum_{k}W_{ik}$. The diffusion equation has a solution 
\begin{equation}
    \vec{u}(t)=e^{-tL}\vec{u}(0)
\end{equation}
with the initial condition $\vec{u}(0)$. 

To quantify the diffusive properties of a given vertex $i$, we study the time scale for the system to approach equilibrium when diffusion starts from $i$, i.e.,  $\vec{u}(0)=\hat{e}_i$. The steady-state projector is
\begin{equation}
    \Pi = \lim_{t\rightarrow\infty} e^{-tL}= \frac{1}{N}\vec{\mathbf{1}}\vec{\mathbf{1}}^T
\end{equation}
where $\vec{\mathbf{1}}$ is the all-one vector. We define the operator 
\begin{equation}
    H(t) = e^{-tL}-\Pi
\end{equation}
Therefore, the deviation from the steady-state solution is
\begin{equation}\label{eqn: h_i(t)}
    h_i(t) \equiv H(t) \vec{u}(0)=\left(e^{-tL}-\Pi\right)\hat{e}_i= \vec{u}(t)-\vec{u}(\infty)
\end{equation}
where $h_i(t)$ is a vector and the subscript $i$ denotes the initial condition $\vec{u}(0)=\hat{e}_i$. Note that $h_i(t)$ decays to 0 in the limit $t\rightarrow\infty$. Similar to the closeness centrality, where the distances to a given vertex are added \cite{Networks_Newman}, we sum over the squared deviations from the equilibrium state of all vertices and define the energy of the system to be 
\begin{equation}\label{eqn: E_i(t)}
    E_i(t)=||h_i(t)||^2=\sum_{j}h_{ji}(t)^2
\end{equation}
which is a strictly decreasing function as $\frac{dE_i}{dt} < 0$. The \textit{mean relaxation time} for the initial condition $u(0)=\hat{e}_i$ is therefore, 
\begin{equation}\label{eqn: relaxation time}
    \bar{T}(i) = \frac{\int_0^{\infty}tE_i(t)dt}{\int_0^{\infty}E_i(t)dt}
\end{equation}
We define the relaxation-time centrality $C_{\mathrm{heat}}$ as
\begin{equation}
    C_{\mathrm{heat}}(i)\equiv\frac{1}{\bar{T}(i)}
\end{equation}
$C_{\mathrm{heat}}$ quantifies the efficiency with which a localized perturbation at vertex $i$ spreads through the network. Vertices with smaller mean relaxation time $\bar{T}(i)$ equilibrate more rapidly and therefore have larger $C_{\mathrm{heat}}$. To investigate the role of suppressed density fluctuations in homogeneous, isotropic disordered networks, we examine the variance of $C_{\mathrm{heat}}$ across vertices.
\subsection*{Calculation of $C_{\mathrm{heat}}(i)$}
We use spectral decomposition to calculate $C_{\mathrm{heat}}(i)$. Since the network is undirected, the Laplacian matrix is symmetric and admits the eigenvalue decomposition $L\phi_k=\lambda_k\phi_k$ with normalization $\phi_k^T\phi_l=\delta_{kl}$. For connected networks, $L$ has a single zero eigenvalue ($\lambda_1=0$) with multiplicity $1$. We label the eigenvalues in ascending order, i.e., $\lambda_1=0<\lambda_2\leq\cdots\leq \lambda_N$. Therefore, we have
\begin{equation}
\begin{split}
H(t) & =e^{-tL}-\Pi=\sum_{k=1}^Ne^{-t\lambda_k}\phi_k\phi_k^T-\frac{\vec{\mathbf{1}}}{\sqrt{N}}\frac{\vec{\mathbf{1}}^T}{\sqrt{N}} \\
& =\sum_{k=2}^Ne^{-t\lambda_k}\phi_k\phi_k^T
\end{split}
\end{equation}
Along with Eq.(\ref{eqn: h_i(t)}) and Eq.(\ref{eqn: E_i(t)}), 
\begin{equation}
E_i(t)= \sum_{k=2}^N e^{-2\lambda_k t}\left(\phi_k(i)\right)^2
\end{equation}
where $\phi_k(i)$ is the $i-$th component of the eigenvector $\phi_k$. It is easy to see that $\frac{dE_i}{dt}=-2\sum_{k=2}^N \left(\phi_k(i)\right)^2 \lambda_k e^{-2\lambda_k t} < 0$, consistent with the physical picture that diffusion is an irreversible process. The spectral decomposition of $E_i(t)$ allows us to evaluate the integrals in Eq.(\ref{eqn: relaxation time}):
\begin{equation}
\begin{split}
    \int_{0}^{\infty}E_i(t)dt & = \sum_{k=2}^N \frac{\left(\phi_k(i)\right)^2}{2\lambda_k} \\
    \int_{0}^{\infty}tE_i(t) dt &  = \sum_{k=2}^N \frac{\left(\phi_k(i)\right)^2}{4\lambda_k^2}
\end{split}
\end{equation}
By making use of the Moore-Penrose inverse (or the pseudoinverse) of the Laplacian matrix \cite{pseudo_inv_ref, pseudo_laplace}, i.e.,
\begin{equation}\label{eqn: L^+}
\begin{split}
    L^{+} & = \sum_{k=2}^{N}\frac{1}{\lambda_k}\phi_k\phi_k^T \\
    \left(L^{+}\right)^2 & = \sum_{k=2}^{N}\frac{1}{\lambda_k^2}\phi_k\phi_k^T
\end{split}
\end{equation}
$C_{\mathrm{heat}}(i)$ can be written in the compact form: 
\begin{equation}\label{eqn: C_heat}
    C_{\mathrm{heat}}(i)= 2\frac{\sum_{k=2}^N\frac{1}{\lambda_k}\phi^2_k(i)}{\sum_{k=2}^N\frac{1}{\lambda_k^2}\phi^2_k(i)} = \frac{2\left(L^{+}\right)_{ii}}{\left(\left(L^{+}\right)^2\right)_{ii}} 
\end{equation}
Therefore, the mean relaxation time of diffusion is dictated by the pseudoinverse of the Laplacian matrix. 
\clearpage
\subsection*{Finite-size robustness of scaling of $R_c(r)$}
\begin{figure*}[h]
\centering
\includegraphics[width=\textwidth]{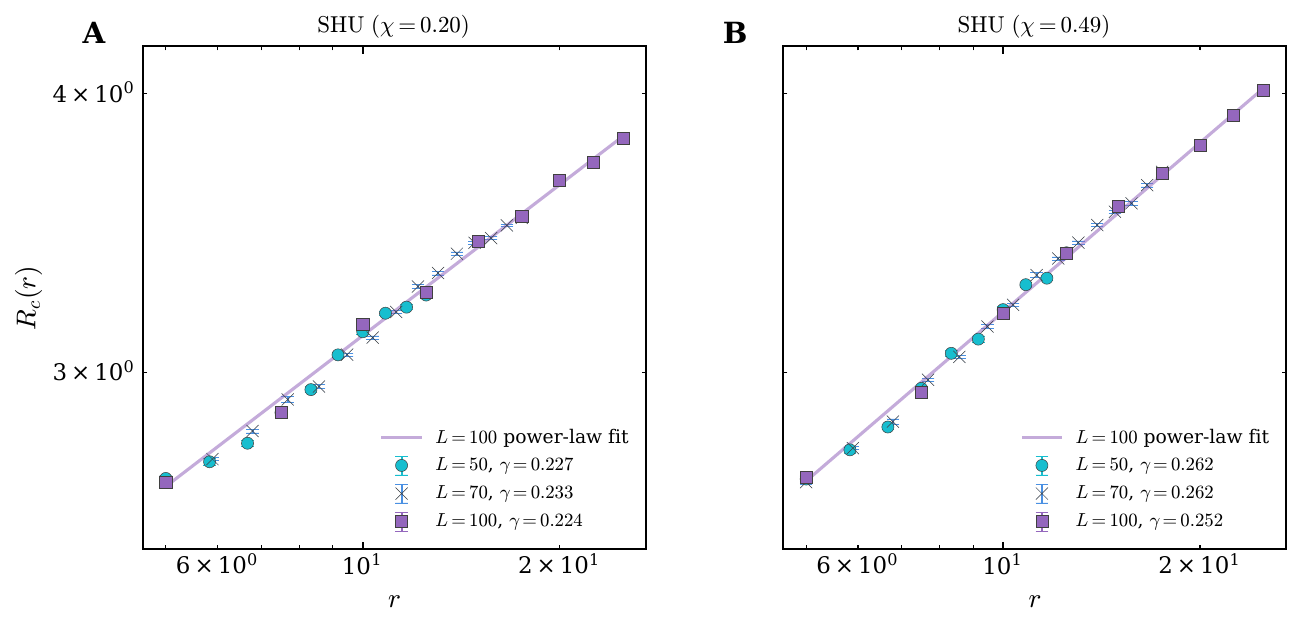}
\caption{Finite-size robustness of the scaling of the critical connection radius $R_c(r)$ with the probed spatial scale $r$ for SHU systems with (A) $\chi=0.20$ and (B) $\chi=0.49$. $R_c(r)$ is well described by the sublinear power law $R_c(r)\sim r^{\gamma}$, with $0<\gamma<1$. The fitted exponents are consistent across the linear system sizes $L=50$, $70$, and $100$, indicating that the observed scaling is robust to finite-size effects over the range examined. The solid lines show the power-law fits for $L=100$, and error bars denote bootstrap standard errors.}
\end{figure*}
\end{document}